\documentstyle[aps,multicol,epsfig,psfig,prl]{revtex}
\tightenlines
\begin{document}

\title{\bf{ Unzipping DNA  - towards the first step of replication }}
\author{ Somendra M. Bhattacharjee\cite{eml1}} 
\address{Institute of Physics, Bhubaneswar 751 005, India}
\maketitle
\widetext
\begin{abstract}
The opening of the Y-fork - the first step of DNA replication - is
shown to be a critical phenomenon under an external force at one of
its ends. From the results of an equivalent delocalization in a
non-hermitian quantum-mechanics problem we show the different scaling
behavior of unzipping and melting.  The resultant long-range critical
features within the unzipped part of Y might play a role in the highly
correlated biochemical functions during replication.
\vskip -1cm
\end{abstract}
\pacs{ 87.14.Gg, 64.90.+b,05.20.-y,03.65.Ge}
\begin{multicols}{2}

DNA, the basic genetic material, is generally a very long, flexible,
linear or circular molecule, with length varying from $2 \mu m$ (5000
base pairs) for simple viruses to $3.5 \times 10^7 \mu m$ ($\sim
10^{11}$ base pairs) for more complex organisms.  In spite of the wide
diversity in the information content and the functionalities of
organisms, the general rules for replication have an astounding
universality in the sense of independence of the system.  A double
helical DNA can be made to melt (i.e. the two strands can be
separated) in vitro (i.e. in the lab) by changes in pH, solvent
conditions and/or temperature (``thermal melting''), but they are
found to be extremely stable in the cell.  As per the standard dogma
of molecular biology, every step of a biochemical process is mediated
by an enzyme (encoded in the DNA), and the process ``that accompanies
DNA replication requires enzymes specialized for this function: DNA
helicases to disrupt the base pairs, topoisomerases to provide a
swivel to unwind the strands, and single-strand binding proteins
(SSB's) to coat and stabilize the unwound strands.  Melting of the
duplex at a replication origin for initiation, and at the fork during
elongation, requires an expenditure of energy, an investment justified
by the functional and evolutionary benefits of maintaining the genome
as a duplex DNA.''\cite{kornberg}

Many biochemical details of the replication process are known,
organism by organism. The proposal of a Y-shaped
structure\cite{levinthal} for a linear molecule\cite{eye} as the
starting point of replication seems to be corroborated by
experiments\cite{kornberg}, but as yet a proper analytical
understanding is lacking.  Very recently, with the advent of various
physical techniques, attempts have been made to study the process from
a physical, rather than a biochemical, point of view.  Special
attention has been given to the measurements of forces to unzip a
double stranded DNA (dsDNA) molecule.  Several studies have been made
to understand the effect of forces in absence of any
enzymes\cite{smith,marko,essevaz,allemand}, while, in other studies,
enzymatic activities involving expenditure of energy and displacement
have been interpreted in terms of effective forces\cite{yin}.  Similar
to ``transcription against applied force'' of Ref. \cite{yin},
experiments combining the mechanical opening of DNA with the
enzymatic replication or transcription have been proposed\cite{marko},
and such an experiment would be highly significant.  The results of
Ref.  \cite{essevaz} distinguish the cases of conventional thermal
melting and the unzipping which has been called {\it directional
  melting}\cite{direct}.  Our analytical approach shows the
differences between these two cases.

Some of the quantitative questions are the following: (1) Is there a
critical force to open up a double-stranded chain in thermal
equilibrium, acting say at one end only? 
(2) Is the nature of the transition different from the
thermal melting of the bound pair and is it reflected in the opened
region? Once these equilibrium questions are understood, we may ask
(3) how the pair opens up in time after the force is applied - the
question of dynamics of unzipping.

\vbox{
\begin{figure}
\begin{center}
\psfig{file=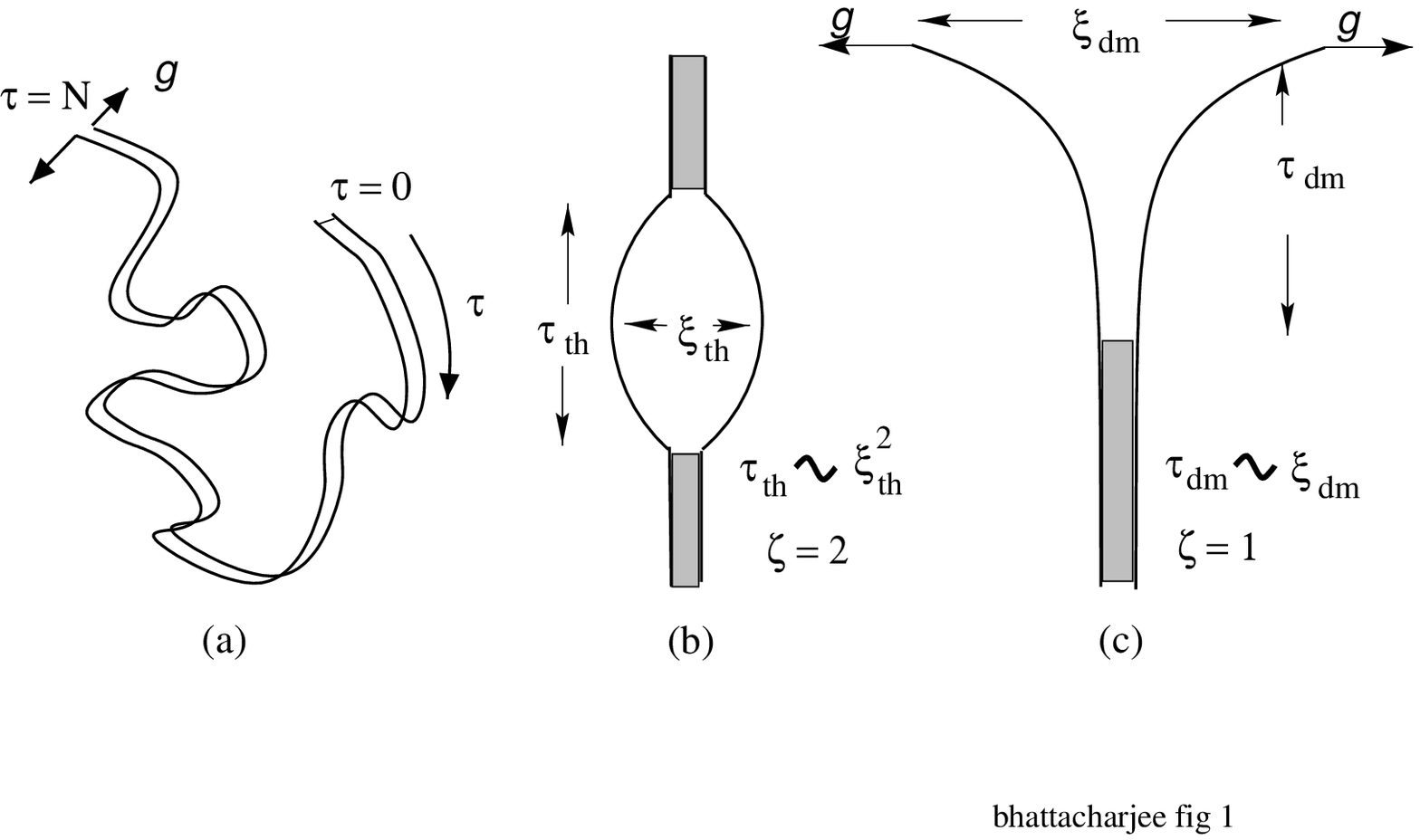, width =3.3in}
\narrowtext
\caption{(a) Pulling the two strands at one end. (b) A ``denatured''
bubble due to thermal fluctuation (``melting''), (c) an unzipped
region slightly below the critical threshold (``directional
melting'').  In (b) and (c), shaded parts denote bound regions.}
\end{center}
\end{figure}
}

In order to focus on the effect of the pulling force (see Fig. 1) on
the bound strands, we take the viewpoint of a minimal model that
transcends microscopic details but on which further details can be
added  for a realistic situation.  Our approach differs from the
previous studies\cite{theory,haijun,cocco} in the emphasis on {\it the
  opening of the fork}.  We treat the DNA as consisting of two
flexible interacting elastic strings.  Winding is ignored.  In a
sense, the effect of topoisomerases is built into the model by
maintaining the chains in the native state.  Such a model has been
found to be useful for many properties of
DNA\cite{pb,wartell,culehwa,pb2,pb3}.  For quantitative results we
consider a bound situation as obtained by a binding square-well
potential.  Our focus is on the simpler aspects of the theory, and
here we answer the first two questions posed above.

Our model is this: Two gaussian polymer chains\cite{doi} in $d=3$
dimensions\cite{lowd}, with $\tau $ denoting the position of a monomer
along the contour of a chain, are tied at one end and pulled by a
stretching force ${\bf g}=g {\hat {\bf e}}_g$ at the other end (see
Fig. 1), ${\hat {\bf e}}_g$ being the unit vector in the direction of the
force.  The energy from the force is proportional to the separation
${\bf r}(N)$ of the end points at $\tau=N$, i.e.
\begin{equation}
-{\bf g} \cdot {\bf r}(N) = -\int_0^N \!d\tau \, {\bf { g}}\cdot
\frac{\partial {\bf r}(\tau)}{\partial\tau},
\label{eq:connec}
\end{equation}
since ${\bf r}(0)={\bf 0}$.  Assuming identical interaction for all
base pairs ( or for a DNA with identical base pairs) the Hamiltonian
can be written in the relative coordinate\cite{mezard}
\begin{eqnarray}
\label{eq:1}
   \frac{H}{k_B T}&=&\int_0^N \! \!\!d\tau \, \left[\frac{1}{2} \left
  (\frac{\partial {\bf r}(\tau)}{\partial\tau}\right
  )^{^{\scriptstyle 2}} - {\bf { g}}\cdot \frac{\partial {\bf
  r}(\tau)}{\partial\tau} \ + V({\bf r}(\tau)) \right ],
\end{eqnarray}
where ${\bf r}(\tau)$ is the $d$-dimensional relative coordinate of
the two chains at the contour length $\tau$. Base pairings require
that the monomers on the two strands interact only if they are at the
same contour length $\tau$.  The potential energy is therefore given
by the integral of the potential $V({\bf r(\tau}))$ over $\tau$.
$V({\bf r})$ is a short-range potential and its detailed form is not
important.  One choice, as generally used for renormalization group
approach, is to take a contact potential $V({\bf r})=v_0
\delta_{\Lambda}({\bf r})$ where, in Fourier space,
$\delta_{\Lambda}({\bf q}) = 1$ for $\mid{\bf q}\mid \leq \Lambda$,
$\Lambda$ being a cut-off reminiscent of the underlying microscopic
structure.  For the equivalent quantum mechanical calculation, we
choose a square-well potential.  
In thermal equilibrium, the properties are obtained from the free
energy $F=-k_B T \ln Z$ where the partition function $Z=\int {\cal DR}
\exp(-H/k_BT)$ sums over all the configurations of the chains.

The Hamiltonian written in the above form can be thought of as a
directed polymer in $d+1$ dimensions about which many results are
known\cite{jj,kolo} for ${\bf g}=0$. If we treat $\tau$ as a time like
co-ordinate, then the same Hamiltonian represents, in the path
integral formulation, a quantum particle in {\it imaginary} time.
This quantum particle with ${\bf g} \neq 0$ then corresponds to the
imaginary vector potential problem much discussed in recent
times\cite{hatano}.  We make use of both the pictures in this paper
and follow the formulation of Ref. \cite{hatano} closely.

For $d=1$, the quantum problem with $V(x)=-v_0\delta(x)$, is exactly
solvable, and is done in Ref. \cite{hatano} as a single impurity
problem.  It was shown that there is a critical $g_c$ below which the
force does not affect the bound-state energy, i.e. the quantum
particle remains localized near the potential well, while for $g> g_c$
the particle delocalizes.  In the polymer picture, this means that in
low dimensions, a force beyond a critical strength separates the two
strands.  It should be pointed out here that the force is applied at
one end only, and but it is not a boundary or edge effect mainly
because of the connectivity of the polymer chain as expressed by
Eq. \ref{eq:connec}.

Details of the phase transition behavior of the Hamiltonian of
Eq. \ref{eq:1} for $g=0$ are known from exact renormalization group
(RG) calculations\cite{jj,kolo} for a $\delta$-function potential.
With $g=0$, there are two fixed points (fp) at (i)$ u^*=0$, and (ii) $
u^*=\epsilon$, with $\epsilon=2-d$ and $u=vL^{\epsilon}$ as the
dimensionless running potential strength parameter, $L$ being an
arbitrary length-scale.  For $d<2$, the first one is unstable and
hence the critical point while for $d>2 (\epsilon <0)$, the second one
is the unstable one.  The unstable fp ($u_{\rm u}^*$) represents the
melting or unbinding of the two chains.  It also follows\cite{kolo}
that other details\cite{intg} of $V({\bf r})$ are irrelevant in the RG
sense, i.e., they vanish in the large length-scale limit, explaining
the universality of the problem.

The important length-scales for this critical point, from the bound
state side, comes from the typical size of the denatured bubbles of
length $\tau_{\rm m}$ along the chain and $\xi_{\rm m}$ in the spatial
extent (Fig. 1b). Close to the critical point these length scales
diverge as $\xi_{\rm m}\sim \mid \Delta u\mid^{-\nu_{\rm m}}$ and
$\tau_{\rm m}\sim\xi_{\rm m}^{\zeta}$, where $\Delta u$ is the
deviation from the critical point, where $\nu_{\rm m}$ and $\zeta$ are
the two important exponents\cite{jj},
\begin{equation}
\nu_{\rm m}=1/\mid d-2\mid \quad {\rm and}\quad  \zeta=2.
\label{eq:melt}
\end{equation} 
It is this $\zeta$ (the dynamic exponent of
the quantum problem) that will distinguish the new phenomena we are
trying to understand\cite{grg}.

The mapping to the imaginary time quantum mechanics allows us to use
the methods of quantum mechanics\cite{hatano}.  As mentioned, we
choose a square-well potential $V({\bf r}) = - V_0$, for $r < r_0$,
and $0$, otherwise.  The quantum Hamiltonian is then given
by\cite{hatano}
\begin{equation}
\label{eq:qham}
H_q({\bf g}) = \frac{1}{2} ({\bf p} + i {\bf g})^2 + V({\bf r}),
\end{equation}
in units of $\hbar=1$ and mass $m=1$, with ${\bf p}$ as momentum.  For
simplicity, the well is chosen to be just deep enough to have only one
bound state (in the quantum mechanical picture, with ${\bf g}=0$) with
energy $E_0<0$.  The non-hermitian Hamiltonian can be connected to the
hermitian Hamiltonian at $g=0$ by
\begin{equation}
U^{-1} H_q({\bf g}) U = H_q({\bf g}=0),\quad {\rm  where}\quad
  U=\exp({\bf g} \cdot {\bf r}).
\end{equation}  
The wave-functions are also related by this $U$-transformation so that
if the transformed bound (i.e. localized) state wave-function remains
normalizable, the bound state energy will not change.  We refer to
Ref. \cite{hatano} for details of the argument.  The continuum part of
the spectrum will have the minimum energy $E=-g^2/2$ (the state with
wave-vector ${\bf k}=0$).  For the localized state at $g=0$, the
wave-function for $r > r_0$ is $\psi_0({\bf r}) \sim \exp(-\kappa r)$
where $\kappa=2{\sqrt{\mid E_0\mid}}$.  The right eigenvector for
$H_q(g)$ is then $\psi_R({\bf r}) \sim \exp({\bf g} \cdot {\bf r} -
\kappa r)$, obtained via the $U$-transformation.  This remains
normalizable if $g < \kappa$, so that the binding energy remains the
same as the $g=0$ value until $g=g_c\equiv\kappa$.  The generic form
of the spectrum is shown in Fig. 2a.  This indicates a delocalization
transition by tuning $g$ - the unzipping or the directional melting of
DNA at $g=g_c$.  The phase diagram is shown schematically in Fig. 2b.
There is a gap in the spectrum (Fig. 2a) for $g<g_c$ and the gap
vanishes continuously as $\mid g^2 - \kappa^2\mid \sim \mid g -
g_c\mid$ as $g \rightarrow g_c-$. Since the time in the quantum
version corresponds to the chain length, the characteristic chain
length for the delocalization transition is therefore\cite{transf}
\begin{equation}
\tau_{\rm dm}(g) \sim \mid g_c-g\mid^{-1}.
\label{eq:tau}
\end{equation}

The spatial length-scale of the localized state is determined by the
width of the wave-function and, for $g=0$, it is set by $\kappa^{-1}$.
For $g\neq 0$, the right wave-function,$\psi_R$, has a different
length scale and this length-scale diverges as the wave-function
becomes non-normalizable.  The width of $\psi_R$ gives this scale as
\begin{equation}
\xi_{{\rm dm}}(g) \sim\mid g_c - g\mid^{-\nu_{\rm dm}}\quad {\rm
 with} \quad \nu_{\rm dm}=1,
\label{eq:xiuz}
\end{equation}
for $g \rightarrow g_c-$.  We find $\tau_{\rm dm}\sim \xi_{{\rm dm}}$
and therefore 
\begin{equation}
\zeta =1.
\end{equation}

\vbox{
\begin{figure}
\begin{center}
\psfig{file=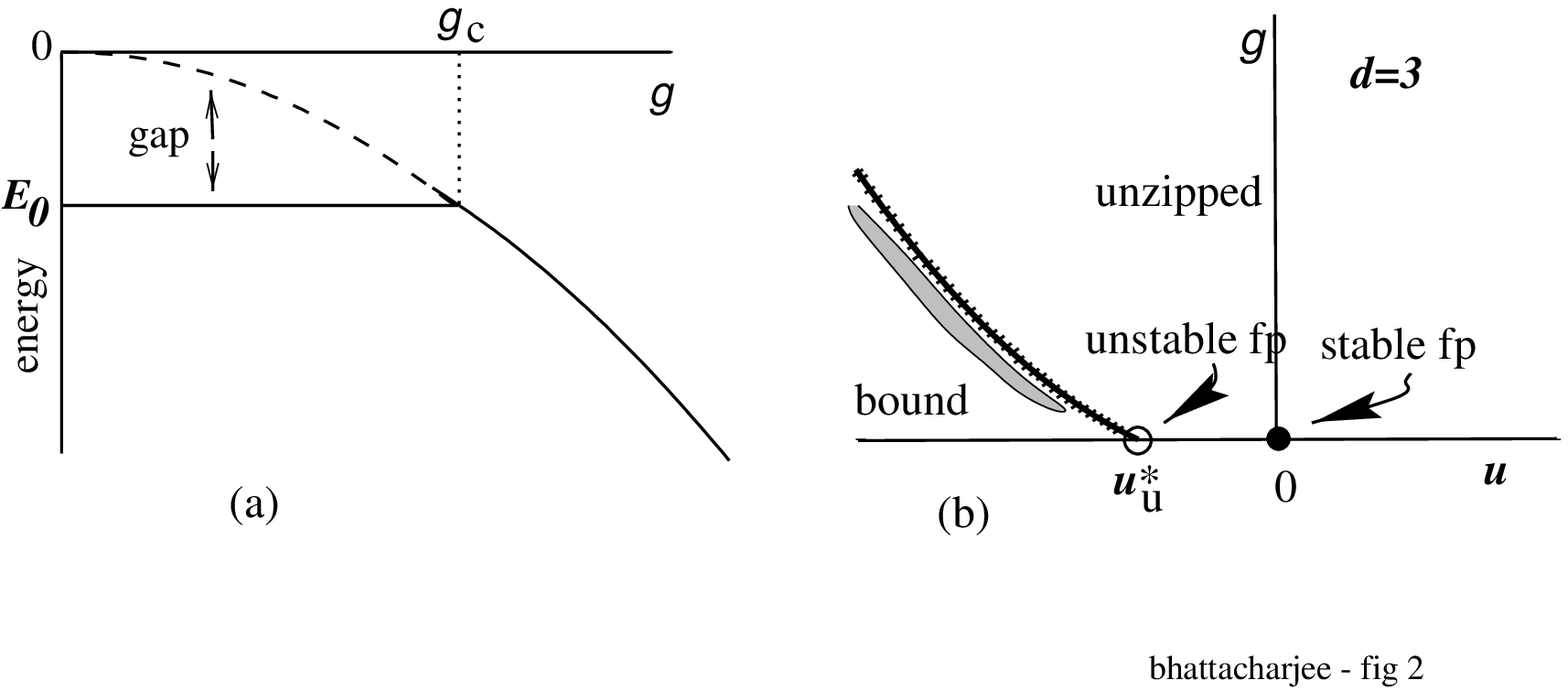, width =3in}
\end{center}
\narrowtext
\caption{ (a)The energy spectrum for $g\neq 0$ for the Hamiltonian in
Eq. \ref{eq:qham}. (b) The phase diagram in the $u-g$ plane.  Thermal
melting takes place along the $g=0$ line at $u=u_u^*$. The hatched
line indicates the unzipping transition or directional melting
line. The biochemical reactions is conjectured to occur at a slight
subcritical region indicated by the hatched region.}
\end{figure}
}

The significance of $\tau_{\rm dm}$ can be understood if we study the
separation of the two chains, i.e., $\langle {\bf r}\rangle_{\tau}$,
at a distance $\tau$ along the chain below the pulled end.  This can
be evaluated by using the standard rules of quantum
mechanics\cite{hatano}. For infinitely long chains in the sub-critical
region, only the bound and the first excited states are sufficient for
the computation.  One finds, along the pulled direction,
\begin{equation}
\langle r\rangle_{\tau} \sim \exp[-\tau/\tau_{\rm dm}(g)],
\label{eq:avx}
\end{equation}
where $\tau_{\rm dm}(g)$ is given by Eq. \ref{eq:tau}.  In other
words, $\tau_{\rm dm}(g)$ and $\xi_{{\rm dm}}(g)$ describe the
unzipped part of the two chains near the pulled end (Fig. 1c).  These
length scales diverge with $\zeta=1$ as the critical force is reached
from below.  (One can also define similar length-scales for $g
\rightarrow g_c+$ where the lengths would describe the bound regions.)
The exponential fall-off, Eq. \ref{eq:avx}, of the separation from the
pulled end immediately gives the picture of the replication Y-fork as
shown in Fig. 1c.

Let us study the behavior of the free energy. Since the partition
function for the Hamiltonian of Eq. \ref{eq:1} obeys a diffusion-like
equation\cite{doi}, the free energy ${\cal F}= F/k_BT - g^2\tau/2$
satisfies the equation
\begin{equation}
\frac{\partial {\cal F}}{\partial \tau} = \frac{1}{2} \nabla^2 {\cal F} -
\frac{1}{2} (\nabla {\cal F})^2 - {\bf g} \cdot \nabla {\cal F}
-v_0\delta_{\Lambda}({\bf r}).
\end{equation}
The left hand side represents the free energy per unit length of the
chains.  Under a scale transformation $x \rightarrow b x$, and
$\tau\rightarrow b^\zeta \tau$, the total free energy remains
invariant so that the above equation takes the form
\begin{eqnarray}
\frac{\partial {\cal F}}{\partial \tau} &=& \frac{1}{2}b^{\zeta
-2}\nabla^2 {\cal F} - \frac{1}{2}b^{\zeta-2}({\nabla {\cal F}})^2
-\nonumber\\
&&b^{\zeta -1}{\bf g}\cdot \nabla {\cal F}- b^{\zeta-d}v_0
\delta_{\Lambda}({\bf r}).
\label{eq:scal}
\end{eqnarray}
For $g=0$, Eq. \ref{eq:scal} tells us that $\zeta=2$ and $d=2$ are
special for melting transition as we see in Eq. \ref{eq:melt}.  For
the choice $\zeta =1$, the $g$-dependent term dominates and all other
terms become irrelevant for large length-scale $b$.  It is this
feature that shows up in the Y-fork of the unzipped chain.  The
robustness of Eqs. \ref{eq:tau}-\ref{eq:avx} also follows from
this. With the dependence on the potential strength entering only
through $g_c$, these are valid along the hatched line of Fig. 2b, and
could be oblivious to the details of the nature of the melting
transition.  In our simple model, the melting point appears as a
multi-critical point in the phase diagram of Fig. 2b.

Now we see the striking difference between the directional melting and
thermal melting (Fig. 1).  In the thermal case the bubbles will have
anisotropic shape with the spatial extent scaling as the square-root
of the length along the chain.  This is the characteristic of the
fixed point $u_{\rm u}^*$, while the pulled case is described by a
different scaling.  The exponential profile of the Y-fork and its
scaling are the two important characteristic features which should be
experimentally verifiable.

We have considered only the equilibrium situation.  The dynamics of
this process of directional melting or unzipping is important.
Similarly the condition of identical interaction may not be realistic
for real DNA.  This feature can be cured by taking the interaction
$v_0$ to be random\cite{rani} with a specific distribution.  Such a
random case\cite{rani} shows a different type of melting behavior.
The mapping to a quantum problem is also then lost.  Self-avoidance
and other topological constraints can also be added to this model,
though at the cost of the simplicity of the model.  For example,
self-avoidance can be introduced in Eq. \ref{eq:1} by adding\cite{doi}
a term $\frac{1}{2}\int d\tau \int d{\tau^{\prime}} v_s \delta({\bf
  r}(\tau) - {\bf r}(\tau^{\prime}))$, with $v_s >0$. The
generalization to a two-chain problem is straight-forward.  Such a
term can be ``unsquared'' by introducing\cite{izyumov} an annealed
gaussian-random potential $V_{\rm I}({\bf r})$ with zero mean and
variance 
$\langle V_{\rm I}({\bf r}) V_{\rm I} ({\bf  r^{\prime}})\rangle$
$ = v_s \delta({\bf r} -{\bf r^{\prime}})$ so that
the equivalent quantum hamiltonian is ${\cal H}_q = H_q + i V_{\rm
  I}({\bf r})$, where $H_q$ is given by Eq. \ref{eq:qham}.  This
involves both an imaginary vector potential and an imaginary, random,
scalar potential.  The polymer problem is recovered from the $V_{\rm
  I}$-averaged propagator of the quantum particle\cite{izyumov}. Such
a general non-hermitian hamiltonian is little understood at present
even for the pure case, let alone the random one.  We wish to come
back to these issues in future.

Let us summarize our results and the emerging picture. There is a
critical strength of the force required to unzip a double stranded
DNA, and this directional melting is a critical phenomenon.  Thermal
or other fluctuations can open up regions along the length of the
chain (bubbles with $\zeta=2$) but the unzipped part is characterized
by a different scaling ($\zeta=1$), as shown in Fig. 1.  In other
words, the Y-fork created by the force represents a correlated region
which is easily distinguishable from a thermal bubble.  Living
organisms probably work at a slightly sub-critical regime with $g <
g_c$ to take advantage of and exploit the distinct correlations for
the enzymatic actions within the Y-fork.  This can be achieved by
coupling the biochemical reactions (like polymerization, unwinding
etc) to the unzipping phenomenon, analogous, in spirit, to
reaction-diffusion systems.  The near-critical features of unzipping
or directional melting then can lead to a coherent phenomenon we see
as replication - a process requiring highly cooperative functioning of
different enzymes in space and time.  This way of viewing the
correlated biochemical events has not been studied so far.  As a first
step, we therefore suggest that high precision measurements be done to
get the profile along the unzipped region of DNA (or simpler
double-stranded polymers) under a pulling force, {\it in vitro} or
{\it in vivo}.

I thank ICTP for warm hospitality, where a major part of this work was
done.  Acknowledgments will probably belittle the influence of a
discussion with Mathula Thangarajh that shaped the final form of this
work.

\vskip -.5cm

\end{multicols} 

\begin{references}
\bibitem{eye} For a circular DNA the starting point is like an
  ``eye'', which might be thought of as two Y's joined together.
\bibitem{direct} Though directional melting is a better nomenclature
  than unzipping, we use both the terms interchangably.
\bibitem{grg} The RG flow equations for the nonzero $g$ case could be
  of the type
\[
L\frac{du}{dL} = \epsilon u - u^2 + c_1 {\tilde g}^2, \ \ {\rm and}\ \ 
L\frac{d{\tilde g}}{dL} = {\tilde g}(1- c_2 u),
\]
where ${\tilde g}= g L$ is the dimensionless force and $c_1,c_2$ are
numerical constants.  These equations do reproduce the $g=0$ flow
equations and have flows that are consistent with the behavior
obtained using the quantum mapping.  This is under investigation.
\end{references}
\end{document}